\documentclass[preprint,12pt]{elsarticle}
\usepackage{graphicx}
\usepackage{lineno}
\usepackage{url}
\usepackage{listings}
\usepackage[colorinlistoftodos]{todonotes}
\usepackage{color}

\definecolor{dkgreen}{rgb}{0,0.6,0}
\definecolor{gray}{rgb}{0.5,0.5,0.5}
\definecolor{mauve}{rgb}{0.58,0,0.82}

\journal{arXiv.org}

    \makeatletter
\def\ps@pprintTitle{%
	\let\@oddhead\@empty
	\let\@evenhead\@empty
	\def\@oddfoot{\reset@font\hfil\thepage\hfil}
	\let\@evenfoot\@oddfoot
}
\makeatother

\begin{document}

\begin{frontmatter}

\title{Service-Oriented Re-engineering of Legacy JEE Applications: Issues and Research Directions}
\author{Hafedh Mili, Ghizlane El-Boussaidi, Anas Shatnawi, Yann-Ga\"el Gu\'eh\'eneuc, Naouel Moha, Jean Privat, Petko Vatlchev }
\address{LATECE Laboratory, Universit\'e du Qu\'ebec \`a Montr\'eal, Canada}

\begin{abstract}
Service-orientation views applications as orchestrations of independent software services that (1) implement functions that are reusable across many applications, (2) can be invoked remotely, and (3) are packaged to decouple potential callers from their implementation technology. As such, it enables organizations to develop quality applications faster than without services. 

Legacy applications are \emph{not} service-oriented. Yet, they implement many reusable functions that could be exposed as \emph{services}. Organizations face three main issues when re-engineering legacy application to (re)use services: (1) to mine their existing applications for reusable functions that can become services, (2) to package those functions into services, and (3) to refactor legacy applications to invoke those services to ease future maintenance.

In this paper, we explore these three issues and propose research directions to address them. We choose to focus on the service-oriented re-engineering of \textit{recent legacy} object-oriented applications, and more specifically, on JEE applications, for several reasons. First, we wanted to focus on architectural challenges, and thus we choose to \textit{not} have to deal with programming language difference between source and target system. We chose JEE applications, in particular, because they embody the range of complexities that one can encounter in recent legacy applications, namely, multi-language systems, multi-tier applications, the reliance on external configuration files, and the reliance on frameworks and container services during runtime. These characteristics pose unique challenges for the three issues mentioned above.  
\end{abstract}


\end{frontmatter}

\newpage
\tableofcontents
\newpage

\section{Introduction}
\label{sec.introduction}

IT departments strain under the needs to automate business processes within, and across, organizational boundaries, and to develop and deploy new applications. They also contend with heterogeneous applications that duplicate some of the same business functions but cannot share most of these functions because of the way that these functions are packaged. Thus, they face the problem of encapsulating functions into reusable services.

Service-orientation purports to solve this problem by viewing applications as orchestrations of independent software services that (1) implement functions that are reusable across many applications or application \emph{domains}, (2) can be distributed and invoked remotely, and (3) are exposed such that their interfaces (a) abstract the specificities of their implementation technology and (b) allow \emph{any client}, regardless of its implementation technology, to invoke them. 

An organization that applies service-oriented development can develop new applications (1) by performing a high-level decomposition of the functions of the applications into coarse functional blocks, (2) by mapping those functional blocks to available services, (3) by implementing the functional blocks that have no corresponding services, and (4) by implementing the process that orchestrates all the services. The more services we have available, the more productive and effective is service-oriented development.

An organization may also want to re-engineer its legacy applications to use service-oriented development (1) by mining the existing applications for reusable functions that could qualify as services, (2) by packaging these functions as \emph{services} to enable their (re)use, and (3) by rewriting some existing applications to (re)use the newly-identified services. Indeed, while service identification and packaging is useful for \emph{future} application development, they do little for existing applications. Therefore, \emph{refactoring} existing applications into service-oriented applications is also important to modernize legacy applications and ease future maintenance.

In this paper, we gather, define, and discuss the most pressing issues and research directions on (1) service identification, (2) service packaging, and (3) the refactoring of legacy applications into service-oriented applications. We survey the literature related to service identification, in particular the many works on identifying services in legacy applications \cite{Huergo:2014:SSS:2663041.2663162,Suwisuthikasem2015,Cai2011}. 

We choose legacy JEE applications as the input of the re-engineering effort for many reasons. First, we chose not to have to deal with \textit{language} or \textit{language paradigm} issues, and focus on \textit{architectural} issues. Thus, we needed a legacy whose programming language(s) can support the SOA style. Second, we chose JEE \textit{in particular}, as opposed to plain Java applications, because JEE applications embody the range of complexities that we are likely to encounter in enterprise applications. Indeed, JEE applications are multi-tier, multi-language, rely on external configuration files to link the pieces together, and rely on frameworks and containers during runtime to connect the pieces together. As such, they illustrate the range of complexities one is likely to encounter in recent legacy enterprise applications. Finally, we chose JEE because the technology has been in use since 1999, and is quite popular for enterprise applications.

This report aims at providing researchers and practitioners with a view on the state of the art on service identification, service packaging, and service-oriented refactoring as well as highlight gaps in the literature. We conclude with a research agenda to discuss with the scientific community and industrial partners.

Section \ref{principles} introduces the principles of service-oriented development and our focus on source code of legacy applications. Section \ref{issues-service-identification} discusses issues related to service identification. We first present a taxonomy of services and then discuss methods for identifying services of each category in the taxonomy. Section \ref{issues-service-packaging} discusses issues related to the packaging of services. Finally, Section \ref{issues-refactoring} describes approaches to refactor the code of legacy applications to (re)use recently identified and packaged services. Section \ref{discussion} discusses the issues and current state of the art while Section \ref{conclusion} concludes with a research agenda.

\section{Principles}
\label{principles}

\subsection{Reusability = Usefulness + Usability}
\label{reusability}

The idea of building new applications from reusable artifacts is not new \cite{Mili1995,Mili2001}. Such artifacts can be analysis-level artifacts (e.g. software models, analysis patterns), design-level artifacts (e.g. architectural styles, design patterns, reference architectures), source-level artifacts (libraries, frameworks), or executables (compiled code, components, services). Reusability has many advertised advantages, including:
\begin{itemize}
	\item \emph{Enhanced productivity}: by reusing existing artifacts of level \emph{i}, we save on development tasks \emph{up to} development stage \emph{i}. By reusing analysis models (or patterns), we save on analysis; by reusing executable components/services, we save on analysis, design, coding, and testing of the functionality implemented by the components/services.
	\item \emph{Enhanced quality}: not only do \emph{domain engineering} methodology suggests that reusable components be thoroughly tested, but re\emph{used} artifacts, whether by design or by accident, will have been tested in \emph{many} different contexts.
	\item \emph{Enforcement of patterns} inherent in the reusable artifacts, be they analysis level patterns, architectural patterns, design patterns, coding styles, etc.
\end{itemize}
We argued in \cite{Mili2001} that \emph{reusability} is a combination of two qualities:
\begin{itemize}
	\item \emph{Usefulness}, which represents the extent to which a reusable artifact can be used in different contexts. A  \emph{sorting} function/procedure or a \emph{collection} data structure are examples of very \emph{useful} utility (domain-independent) artifacts.
	\item \emph{Usability}, which represents the extent to which the functionality is packaged in a way that facilitates its use in those contexts where it is \emph{useful}. If I am programming in Java, a C sort routine is of little use to me.
\end{itemize}
Our work on service-oriented reengineering will deal with both issues:
\begin{itemize}
	\item Service identification will deal with issues of \emph{usefulness}: identify those clusters of functionality that are used/invoked in \emph{many} places within the same application, or across \emph{many} different applications.
	\item Service packaging deals with \emph{usability}, as it attempts to wrap/package that reusable functionality behind \emph{service interfaces}, which will facilitate their reuse (remoteness, technology independence, etc.).
\end{itemize}
\subsection{It is architectural style migration}
\label{style-migration}
The problem that we are trying to solve, i.e. the \emph{service-oriented reengineering of legacy JEE applications}, is an instance of the more general problem of \emph{architectural style migration}: how to migrate an application from an architectural style \emph{source} to an architectural style \emph{destination}, \emph{while keeping other things equal}. While migrating legacy COBOL applications towards a Java-based service-oriented applications may be of utmost practical relevance, such an endeavor requires dealing with \emph{many} differences between the source and target technologies:
\begin{itemize}
	\item Dealing with different programming \emph{languages} that are generations apart, at the programming language \emph{construct} level (typing, procedural invocation, parameter passing, control structures, data reference).
	\item Notwithstanding statement/construct-level difference, we would have to deal with different programming \emph{paradigms} (as in going from C to C++).
	\item Dealing with different virtual--and physical--machine models (e.g., how processes are run, how concurrent access to shared resources is done,etc.).
	\item And, \emph{also}, dealing with different \emph{architectural styles}.
\end{itemize}
Because we want to focus on the issues dealing with \emph{architectural style migration}, apart from the other issues, we chose a recent legacy--in this case a JEE web application--so that we don't have to deal with language or \emph{programming paradigm} issues. Thus both the source and target application will be written in Java\footnote{\emph{Mostly}, in addition to all the other scripting and 'data specification' languages, such as XML-based configuration files}. Similar work can be explored with recent legacy in other languages (e.g. C\#). While the \emph{tools} that we develop may not be portable across languages, the issues--and solution approaches--should be similar.

\subsection{Focus on source code}
\label{source-code-only}

Service identification is a difficult problem, in part because there are different kinds of services, which have different characteristics (see section \ref{service-taxonomy}), and in part because there are domain-specific considerations that may determine what makes a 'good service'. The literature on service identification uses a number of techniques that often combine \emph{many} knowledge sources, \emph{including} source code, but also including domain knowledge, as provided by subject matter/domain experts, or \emph{domain-}/\emph{analysis-level} models, that provide a more abstract representation of the functionality of the software [references]. Other work combines \emph{static} code analysis with \emph{dynamic} code analysis, that looks at execution traces of the legacy software at hand, which gives a better idea about the \emph{effective/actual usefulness} of a piece of code (how often it \emph{is used}), as opposed to the \emph{potential usefulness} of a piece of code, as established by program dependency graphs, for example\footnote{The \emph{presence} of a call relationship between two methods indicates a \emph{possibility}, depending on whether the control path along which the call is made is actually taken during execution}.

For this work, we choose to focus on \emph{static analysis of source code}, for several reasons:
\begin{itemize}
	\item Dynamic code analysis (run-time tracing) is \emph{usually} fairly complicated to set-up, and the techniques that we develop can only be applied to relatively 'small' and 'simple' legacy applications. For example, in a client-server application, both the client and the server need to be traced. When the two do not run on the same machine, it becomes difficult to collect-- and correlate-- the traces.
	\item Domain or analysis models, which are sometimes used to complement static code analysis (reference) may not always be available. And when they are, they are \emph{most likely} out of sync with the source code.
	\item We are not so much interested in advancing the state of the art in \emph{service identification}, which has received \emph{considerable attention} in the past decade or so, as we are interested in issues dealing with \emph{service packaging} and \emph{service refactoring}, which, in our view, did not get the attention they deserve. The technical challenges involved in service packaging and refactoring are, as far as we can tell, mostly independent of the \emph{intrinsic} quality (the \emph{usefulness}) of the services. By foregoing those additional knowledge sources (execution traces or domain/analysis knowledge), we may get lower quality services, but that does not prevent us from, or affect our ability to, explore the issues related to packaging and refactoring.
\end{itemize}
We do not exclude input from human experts to annotate/qualify intermediate or final results of service identification to enhance the quality of the services, but we will refrain from making such input a \emph{necessary} step in service identification.

\subsection{KDM as a repository for architectural level knowledge}
\label{use-KDM}
Service identification, repackaging, and refactoring require that we manipulate representations of the legacy application at various levels of abstraction, going from instruction-level representation to detect method/procedure invocations or perform data flow analysis, to representations of \emph{software artifacts} including files of different types (HTML, JSP, JSF, Java, various dialects of XML, property files, etc.), to representations of architectural-level concepts such as the notion of \emph{package}, \emph{layer}, \emph{interface}, and eventually, \emph{service}.

Because of the interdependencies between the different representations-- and reengineering tasks-- we need a representation that:
\begin{itemize}
	\item \emph{integrates} the different views/representations of the legacy application, and
	\item we can share with researchers working on similar problems, or use to compare our work/algorithms to those of other researchers.
\end{itemize}
To this end, we propose to use OMG's \emph{Knowledge Discovery Metamodel}, which provides a '... common \emph{intermediate representation} for existing software systems and their operating environments' (http://www.omg.org/technology/kdm/). The main page of OMG's KDM standard further defines KDM as:
\begin{quote}
KDM is a metamodel for knowledge discovery in software. It defines a common vocabulary of knowledge related to software engineering artifacts, regardless of the implementation programming language and runtime platform - a checklist of items that a software mining tool should discover and a software analysis tool can use.	
\end{quote}
Specifically, we will use the MODISCO (partial) implementation of KDM, which is an open-source Eclipse project (http://www.eclipse.org/MoDisco). MODISCO, available as an Eclipse plug-in, includes a number of tools that can parse files of different types and build KDM representations of the contents of those files. Within the context of our work, we expect to write our own parsers/builders to analyze the different file types that we are likely to find in a JEE legacy application, and build the corresponding KDM representations. Such parsers/builders may include Eclipse's own JDT (Java Development Toolkit), which knows how to analyse/parse Java projects.
\section{Issues in Service Identification} \label{issues-service-identification}
\subsection{What is a service}
\label{what-is-a-service}
\begin{figure}[h]
\centering
\includegraphics[width= 0.9\linewidth]{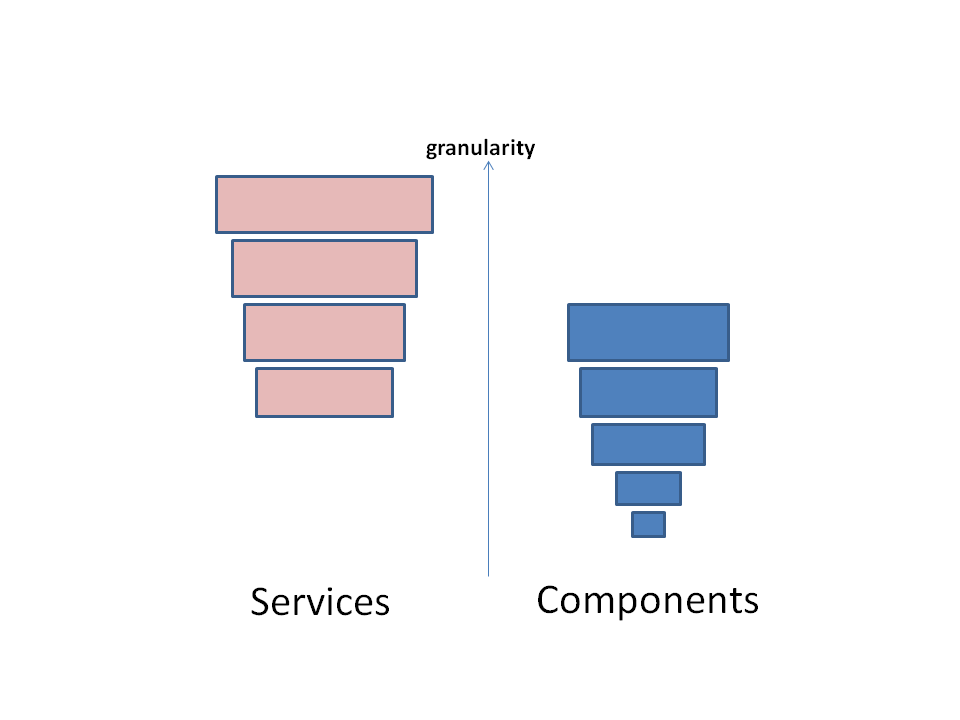}
\caption{Services can be fairly coarse-grained; components can be fairly fine-grained}
\label{fig.services-vs-components-granularity}
\end{figure}

There has been much debate in the community about what constitutes a service. Thomas Erl (Reference), an SOA pioneer, identified eight characteristics of services :
\begin{itemize}
	\item \emph{Standardized [service] contracts}. As software components, services define their capabilities using a standard, implementation-neutral language.
	\item \emph{Loose coupling}. The services are loosely coupled, and any dependencies are explicitly stated in their service contracts.
	\item \emph{Abstraction}. Whereas loose coupling refers to dependencies between services, abstraction refers to dependencies between a service provider and a service consumer. The consumer should not depend on the implementation details of the service.
	\item \emph{Reusability}. Services embody reusable functionality that can service many consumers. In other work, we defined reusability as usefulness and usability (Reference). Usefulness refers to how often the provided functionality is needed while usability refers to how easy it is to use. Usability embodies many aspects, including the existence of (standardized) service contracts (see above), as well as discoverability, composability, and interoperability, discussed below.
	\item \emph{Autonomy}. From the perspective of the consumer, services should be perceived as self-contained components with total control over their resources and environment. The consumer should be able to assume that the service needs no more than the parameters specified in its service contract to do its job. Naturally, behind the scenes, a service may in turn depend on other services. For example, business services can depend on a layer of shared technical services.
	\item \emph{Statelessness}. We can understand statelessness of services in two complementary ways. To be able to 'service' many consumers, a service should not have to rely on implicit state information about its consumers; all of the data needed to service a particular consumer's request should be explicitly passed as parameter. The second aspect of statelessness is related to multiple interactions with the same consumer. This means that a consumer can invoke the operations of the service as many times as they want, in any order they wish, and always get the same result.  In practice, of course, these two conditions are seldom attainable-and not necessarily desirable. If I am using a flight booking service, I sure hope that my interactions with the service have a lasting effect on the state of the world: the creation of a booking in the booking database. Erl writes "Applying the principle of service statelessness requires that measures of realistically attainable statelessness be assessed, based on the adequacy of the surrounding technology architecture to provide state management delegation and deferral options" (see footnote 17).
	\item \emph{Discoverability}. This refers to the ability of services to document and advertise their capabilities so that service consumers can find them. The documentation of the capabilities of a service needs to be expressed in a domain language that is distinct from the language used to express the service contract.
	\item \emph{Composability}. This refers to dual capability of services to, a) be composed at arbitrary levels of aggregation to form more complex services, and b) address many needs. This, in turn, influences two design aspects of services, a) the modalities for interacting with the service, and b) the way the capabilities of the service are distributed among its operations.
\end{itemize}
All of the characteristics, but for \emph{discoverability}\footnote{One might want to distinguish between \emph{definitional properties/characteristics} of services, which \emph{all} services must have, from \emph{characteristic properties/characteristics}, which \emph{most} services will have but which are not necessary. We could also distinsguish between content-related characteristics, and packaging-related characteristics. Within the context of \emph{web services}, discoverability relates to the extent to which a description of the service functionality is available and published in a service registry \emph{\`a la} UDDI.} apply, to a large extent, to \emph{software components}. There has been much debate in the literature about the difference between services and components. We see three differences:
\begin{itemize}
	\item {Packaging}. With \emph{services}, there are existing \emph{standards} that prescribe, a) how services expose their functionalities (e.g. using WSDL files, for the case of web services, or SCA interfaces, for SCA), and b) how their functionalities are invoked (e.g. using the HTTP/SOAP or REST protocols). However, no such standardized protocols exist for \emph{components}\footnote{There are, of course, \emph{component models}, such as Microsoft's COM/DCOM models, or Java-based EJBs, however, they are language/technology/vendor-specific.}.
	\item {Lifecycle}. \emph{Typically}, components are 'dormant' when not needed. They come to life when solicited, service the incoming request, and die out upon termination; in other words, their lifecycle is managed, directly or indirectly, by their client applications ('consumers')\footnote{The lifecycle of components, within a particular technology, may be managed by \emph{containers} for that technology, as is the case for the Windows operating system, for COM components, and for JEE servers, for the case of EJB entities}. By contrast, services \emph{tend to be mostly} independently running processes/services, with their lifecycle managed independently from their client applications.
	\item {Granularity}. Components can be as small as an ActiveX control (a button, or a selection-in-list widget), while services can be as big as a 'business process' (see Section \ref{service-taxonomy}). This is perhaps a corollary of the previous two characteristics: the overhead involved in the packaging and lifecycle management of services makes it ineffective for small-grain functionality-- although, \emph{microservices} tend to blur the lines.
\end{itemize}
Figure \ref{fig.services-vs-components-granularity} illustrates the difference in granularity between services and components.

We suspect that, with the exception of granularity, the differences between services and components will not matter for \emph{service identification}, as they deal mostly with packaging and deployment of the identified services, but not with their functional content, or patterns of usage, per se. More on the criteria for service identification in section \ref{service-identification-approach}.

\subsection{A service taxonomy} 
\label{service-taxonomy}
Many authors/standards pointed out that it is important to define and use a service taxonomy when implementing an SOA \cite{Erl2007,Shy2007,OpenGroup}. Many service taxonomies were proposed in the literature (e.g., \cite{Erl2007,Shy2007,Fuhr2011,Alahmari2010}). These taxonomies tend to classify services according to a hierarchical layered scheme that mainly supports communication between stakeholders during the implementation of SOA initiatives. Most of these taxonomies classify services according to their granularity (e.g., \cite{Erl2007,Shy2007,Alahmari2010}) and/or their reuse (e.g., \cite{Shy2007,Fuhr2011}).

In the Opengroup SOA reference architecture (RA), services are classified according to the capabilities they are providing \cite{OpenGroup}. The Opengroup taxonomy describes a broad functional categorization scheme which includes: strategy and planning services, management services, development services, business services, infrastructure services, asset and registry services, interaction services, process services, information services, access services, partner services, business application services and life-cycle services. Most of the other taxonomies (e.g., \cite{Erl2007,Shy2007}) concentrate only on the core services implementing an SOA solution, i.e., services implementing the business processes, domain-specific functions, and infrastructure/utility functions. 

All proposed taxonomies distinguish between domain-specific services and domain-neutral (or generic) services. Domain-specific services, commonly referred to as business/application services represent/implement business processes and domain-specific functions. Depending on their granularity and scope of reuse, most of existing taxonomies identify the following types of domain-specific services:
\begin{itemize}
	\item Business services: Named process or task services in some taxonomies (e.g., \cite{Erl2007,Shy2007}), they correspond to business processes or use cases. These are the services that are generally consumed by the end-users; for instance a service that enables an on-line booking. To implement business processes, these services compose the capabilities/functions provided by the Enterprise, Application and Entity services described below.
	\item Enterprise services: Named capabilities in \cite{Shy2007}, they are of finer granularity than Business services. They implement generic business functions that are generally reused across different applications; for instance a service that computes taxes.
   	\item Application services: These services implement business functions that are specific to an application. They may be created to support reuse within the application scope or to enable the decomposition of a complex business process \cite{Shy2007}. Similar to Enterprise services, these services are of finer granularity than Business services.
	\item Entity services: They are also called information or data services depending on the taxonomy. They provide access to and management of the persistent data of the business. They generally support actions on data (create, read, update and delete), and may have side-effects (i.e., they modify shared data).
\end{itemize}

In their broadest definition, domain-neutral services provide common facilities that enable using, managing, developing and integrating domain-specific services. We distinguish two categories of these technical services:
\begin{itemize}
	\item Utility services: They may be seen as services that do not support directly the business processes of the company. They generally embody some cross-cutting capabilities required by domain-specific services. Logging and security services are examples of such services.
	\item Infrastructure services: These services represent the facilities that are required for deploying and running an SOA application. Examples of such services are integration services which include services for routing, protocol conversion, message processing and transformation. These services are typically supported by an Enterprise Service Bus (ESB). Compared to utility services, infrastructure services have a broader scope of reuse.
\end{itemize}

Considering the different categories of services identified above, we propose a multidimensional taxonomy that aims at supporting the identification of services from the legacy source code. The dimensions of this taxonomy are:
\begin{itemize}
	\item Domain: domain-specific (business) versus domain-neutral (technical).
	\item Granularity: fine-grained versus coarse-grained. The granularity of a service depends on the complexity of the capability/function the service provides.
   	\item Scope of reuse: Enterprise versus Application. This dimension is dependent on the two previous ones: domain and granularity. Fine-grained services are likely to be more shareable as they may be composed to provide different coarse-grained services that may not belong to the same application. Likewise domain-neutral services, in particular infrastructure services, may be reused across different applications.
	\item Side-effects: computation-only services versus services with side-effects. Services with side-effects are those that manage the application state; i.e. they modify the persistent data of the application. These are mainly Entity services. To maintain data consistency, these services require the implementation of some transactional and/or compensation mechanisms.
    \item Visibility: This refers to the fact that end-users/external services/partner services interact directly with the service or not. Coarse-grained domain-specific services (i.e., Business services) are generally those that are visible.   	
\end{itemize}

Table \ref{table.service-taxonomy} summarizes the properties of the different categories of services according to the dimensions of our taxonomy. In the context of a legacy to SOA migration process, service categories and their properties must be taken into consideration while building a service identification approach (see discussion in  Section \ref{service-identification-approach}). 
 
\begin{table}[ht]
	\centering
		\begin{tabular} {p {2.5 cm} p {1.7 cm} p {2 cm} p {2 cm} p {2 cm} p {2 cm}}
		\hline \hline \\
		Service category & Domain & Granularity & Scope of reuse & Side-effect & Visibility \\ 
		\hline \hline \\
		Business & Specific & Coarse & Application & No & Yes \\ \hline \\
		Enterprise & Specific & Variable & Enterprise & No & No  \\ \hline \\
		Application & Specific & Variable & Application & No & No  \\ \hline \\
   		Entity & Specific & Fine & Variable & Yes & No \\ \hline \\
   		Utility & Neutral & Variable & Variable & No & No \\ \hline \\
   		Infrastructure & Neutral & Variable & Enterprise & No & No \\        
		\hline
		\end{tabular}
	\caption{A multidimensional service taxonomy}
	\label{table.service-taxonomy}
\end{table}

Beyond the fact that different service types require different identification strategies, the most appropriate SOA deployment technology (e.g., Web services versus REST), may depend on the service type. While REST seems to be a very good fit for implementing Entity services with its uniform interface, it's not granted that it is a good fit for the other types of services. Many factors, including the needs for more advanced caching features and security requirements than those provided by REST, makes it worth investigating the features supported by SOAP\footnote{REST relies on the HTTP caching and security features. In specific contexts (e.g., Stock exchange), dynamic resources will require a more appropriate caching than that provided by HTTP. Also, HTTP provides  a very limited support to security when compared to WS-Security the SOAP specification for applying security to Web services.}. 

\subsection{Service identification: a (brief) survey} 
\label{service-identification-survey}
Generally existing service identification methods (SIMs) use different techniques to identify services. These techniques include model-driven approaches, clustering and architecture recovery  techniques and ontology mappings.
	
Alahmari et al. \cite{Alahmari2010} propose a model-based SIM. They propose a service classification that is based on the granularity of the services where the granularity of a service is determined by the number of messages and the complexity of data exchanged. To identify services, they propose a hybrid model-driven approach. They manually build a knowledge portfolio that contains analysis models describing main functions and components of the system. This is done using questionnaires, interviews and available documents. They also derive UML class diagrams from legacy code. Relying on the business functions defined in the knowledge portfolio, they manually build activity diagrams from these class diagrams. These activity diagrams are transformed automatically into BPMN models which are used to identify atomic processes and related business entities. The identified processes are stored in a process portfolio. Candidate services are identified from the process portfolio using metrics and rules that gather/partition atomic processes and their related entities. These metrics and rules are not detailed. The candidate services are then classified and evaluated according to an SOA meta-model and a set of rules to generate services with an optimal level of granularity.

Zhang et al. \cite{Zhang2005e} propose an architecture-based SIM. The proposed approach starts by recovering design and architecture information using static and dynamic analysis. The extracted information is documented using an architecture description language. Service identification is carried out in two steps. The first step consists of analyzing the domain and building a domain model. The model is then used to identify the business functionalities that need to be provided as services, called logical services. In the second step, the recovered architecture information is analyzed to identify legacy components. This is done using hierarchal clustering techniques and modularization criteria (e.g., coupling, cohesion). A legacy component is then matched to the logical service that must offer the functionalities embedded in the component. Legacy components are integrated with newly built components and packaged into Web services.

Greiger et al. \cite{Grieger2014} proposed an approach whose focus is to identify composite services which are the entry points for end users. The approach does not really identify other finer services. In particular, an initial service design is built by mapping each single module to a business service as defined in the classification in \cite{Alahmari2010}; i.e. one of enterprise, application or entity service, according to our taxonomy. A composite service is introduced to aggregate and orchestrate these services. This initial service design is improved through a two-step process. In the first step, hierarchical clustering is applied to the business services depending on the navigation flow between them. Composite services are then introduced to aggregate the resulting clusters of business services. Hierarchical clustering is performed in an iterative way: at each iteration, new clusters are computed for a hierarchal layer and composite services are introduced to realize these clusters.  In the second step, the functionalities of several services are partitioned into parts and common parts are moved into a single service. This partitioning is based on detection and removal of software clones.

Service identification methods were studied in a number of surveys (e.g., \cite{Gu2010,Cai2011,Huergo2014}). We briefly present some of these surveys in the following subsection.

\subsubsection{Overview of existing SIM surveys} 
With the goal of supporting practitioners in selecting a service identification method (SIM), Gu and Lago \cite{Gu2010} carried out a systematic review of 30 SIMs. The review focused on three critical aspects of SIMs: the inputs to these methods; the type and the “format” of the services they produce; and the strategies and techniques they use. The authors identified seven types of inputs including business processes, application domain (i.e., models or documents that describe different aspects of the domain) and legacy system. According to the results of this survey, most of existing SIMs adopt a top-down approach when implementing SOA and take as input business processes and domain information. Bottom-up SIMs rely on source code and its architecture to identify services. Very few methods adopt a hybrid approach that combines legacy systems and other types of inputs. Regarding the service type, most of existing SIMs identify business services and few target technical services. Most of the SIMs describe the identified services in an informal way using a list of terms (e.g., description, input, output). To identify services, the primary strategies used by the studied SIMs involve the decomposition of business processes or business function models. These strategies are combined with different techniques including formal rules encoded in the form of algorithms and less formal techniques provided as guidelines.

Cai et al. \cite{Cai2011} surveyed a number of SIMs with the goal of highlighting their shared high-value activities and practices. In particular, they study the relation between service identification and the SOA engineering process (i.e., forward engineering vs re-engineering). In a forward engineering process, service identification is part of the requirement specification and analysis processes and it includes defining, analyzing, refining and modeling functional and QoS requirements. In a re-engineering process, service identification is part of the reverse-engineering stage and it involves analyzing both code and documents of the existing application, wrapping existing modules and consolidating dispersed functions. Though SIMs adopt different processes, quite a few activities, called high-value activities, are shared among these methods. Cai et al. argue that selecting and composing these activities may be an effective way to identify services. They identify model-driven and decomposition activities as high-value activities in top-down SIMs while identification of reusable legacy assets is the most frequently used activity in bottom-up SIMs.

Most of existing surveys focus on the service type identified by existing approaches. Some of these surveys consider the techniques used by these approaches, the input they require and the output they produce. Very few studies considered how these approaches manage the granularity of the identified services. None of the surveys analyzed the impact of the SOA targeted technology on the identification method.

\subsection{Service identification: our strategy}
\label{service-identification-approach}

Most of the works on service (and component) identification described above relies on the modularity of a service or component---in addition to domain semantics---as reflected in its \emph{functional cohesion}, i.e., the extent to which the different parts of a service or component are related to the same functions, and its \emph{coupling} with the outside world, i.e., the extent to which it depends on other services or components to perform its functions. Thus, several of the approaches rely on \emph{code coupling metrics} that measure code-level dependencies between lower-level constructs, such as \emph{classes}, which are then used by \emph{clustering algorithms} where resulting clusters are proposed as candidate components/services.

In light of the different kinds of services described in Section \ref{service-taxonomy}, we believe that cohesion and coupling \emph{are not sufficient} to characterize services, among all potential clusters of functions. They may \emph{even be unnecessary}. Take the example of a \emph{security service} that stands between an application and its clients, checking whether a particular client has the right to invoke a particular function. We can have two programming models:

\begin{enumerate}
\item Clients request the execution of functions by talking to the security manager, who checks their credentials against a stored security policy, and  decides whether to forward the request or not.

\item Clients talk directly to the server public API, who first double-check with the security manager, before proceeding with the requested call.
\end{enumerate}

In the first case, the security service has a strong coupling with both clients and servers: it is called by \emph{all} clients and it calls \emph{all} server functions. In the second case, the security manager is called by \emph{all} server functions. Either way, the functions offered by the security manager would not appear \emph{alone} within its own dependency cluster and it would not be proposed as a candidate service.

Thus, any service identification approach must take into account the specific types of services that they identify. They must use the metrics that are appropriate for that types. More generally, \emph{component identification approaches for a target architectural style must take into account the specifics of the components types for that style}. As a first step, we must identify, for each service type, the kind of signature that such a service type have in the code \emph{and then} develop approaches to detect such signatures. We believe that not all service types have distinct \emph{code smells} and two different service types may leave similar or indistinguishable signatures in the code.

Further, to go back to our security manager example, we may identify such interceptor-type service by looking for classes that have both high fan-in (everybody calls them) and high fan-outs (they call everybody). However, we thus assume that the functions of the security manager have been packaged within the same class. For example, the security manager may have distinct methods for checking whether a caller can (a) create an object of a particular type, (b) read the value of objects of a certain type (or certain objects), (c) modify objects of a particular type, or (d) delete objects of a certain type. However, if those methods are not part of the same class, we face two additional difficulties:

\begin{itemize}
\item The code signature (high fan-in, high fan-out) may not be as visible.

\item The four independent functions are distinct but should be in the same service.
\end{itemize}

Thus, our service identification algorithms need to rely on code smells of \emph{service functionality that is not yet packaged as a service}.

Table \ref{table.service-identification-criteria} shows a preliminary mapping between service types and service identification criteria. \textbf{Not all pieces of code (classes, methods, etc.) belong to services, some may belong to non-reusable, non-visible, non-service ``thing'' (where a ``thing'' could be a library or utility service.}

\begin{table}
\centering
\begin{tabular} {p {2 cm} p {2 cm} p {2 cm} p {2 cm} p {2 cm} p {2 cm}}
\hline \hline \\
Service type & Examples & Characteristics & Code signature & How to recognize & related work \\ [0.5ex] \hline \\

Utility service & A security service, a logging service & Implements a cross-cutting function & Implements an interceptor-like or filter-like pattern & High fan-in and fan-out back into the application (as opposed to infrastructure, see below) & Works on aspect mining such as \cite{Tonella2004,Robillard2002,Marin2006}

\\

User-defined ``Infra\-structure'' service & A service that uses the underlying infrastructure & Pervasive services & Low fan-out into the application, high fan-out into the infrastructure (known APIs) & Fan-in from higher layer(s), fan-out to infrastructure & Extracting layers in layered architectures (e.g., \cite{ElBoussaidi2012,Belle2014}) 

\\

Entity service & Creating an order, retrieving payments made on an order &  CRUD operations on \textit{configurations} of \textit{domain objects} & Different patterns, depending on C, R, U, or D & See below & Criteria for REST services might be useful 

\\

Enterprise service & Computing taxes, scoring applications, rendering decisions & Supposed to have no side effects. See below & Similar to entity services, except for side effects & Same as with entity services, except side effects & Not sure

\\

Application services & Etc. & Smaller granularity than business services. See below & Complexity, combination of more elementary services & May not be distinguishable from business services & Unknown 

\\
        
Business services & Booking a flight, making an on-line purchase & they implement a business process/use case & complex orchestrations of smaller domain services& there is not a single pattern to look for & Existing work relies on external knowledge such as BP models 

\\

\hline
\end{tabular}
\caption{Service type identification criteria: preliminary proposal}
\label{table.service-identification-criteria}
\end{table}

Let us elaborate on some of the cells of the table:

\begin{itemize}
\item With regard to \textit{entity services}, the way to detect their occurrences depends on the kind of operation: C, R, U, or D. For C and U, concentrated use of constructors, setters, and more generally, methods returning void, for R, use of getters on individual objects or configurations of objects (e.g., navigating associations). 

\item With regard to \textit{enterprise services}:
\begin{itemize}
\item regarding their characteristics, they are supposed to have no side-effects. However, when the computations are complex, we may need complex objects to return them, thus they may have side-effects in the \textit{programming sense}, but do not result into modifying persistent data.

\item Regarding their code signatures, they may follow similar patterns to an R pattern for an entity service, with significant computation (perhaps more than simple aggregation) that involves external data/constants.

\item Regarding ways to uncover them: the best way to test their (re)use across applications \textit{is to have several applications}: if we had a portfolio of enterprise applications, then we can look for reusable clusters, which could be enterprise services.
\end{itemize}

\item With regard to \textit{application services}:
\begin{itemize}
\item Regarding their characteristics, although they represent orchestrations of other services (e.g., entity or enterprise services), they are application-specific, reusable within application, but not across applications.

\item regarding ways to uncover them, they will exhibit similar levels of complexity to business services in terms of error handling, multi-threading/asynchrony, compensation, and so forth, and, thus, they will not be distinguishable from business services. However, we probably do not care about the distinction, which is mostly granularity and reuse scope
\end{itemize}

\item With regard to \textit{business services}:
\begin{itemize}
\item Concerning their code smells, how do we detect/guess that a piece of code implements a business process? Their functionalities may exhibit a significant error handling or compensation component. They are transactional, possibly involving different databases or servers. They typically result into many changes to disjoint databases (multi-party transactions), multi-threading, asynchronous execution, etc.

\item Regarding ways to detect them, there is no single pattern: it is a combination of evidences for each of the previous characteristics (transaction frontiers, use of messaging, complex error handling, etc.)
\end{itemize}
\end{itemize}

This table provided some fairly preliminary hypotheses. More thought needs to go into the different categories. Also, we must decide whether the distinction between different types of services actually matters.

In the next section, we discuss the issues raised by the static analysis of legacy JEE applications.

\subsection{Issues in static code analysis of legacy JEE Application}
\label{issues-static-code-analysis}

Having decided to perform service identification using source code as our only input, we now look at some of the issues raised by analyzing the source code of JEE applications. We first present an overview of the issues, then discuss some of the issue is some detail.

Static code analysis relies on a set of graph-like structures that represent the relationships between program elements. Such structures provide the basis for computing the various metrics that we will need for our service identification algorithms. As a preliminary step to service identification, we need to compute such structures in a way that identifies \emph{all} of the relationships that exist between elements, \emph{despite} the different language, technology, and development environment mechanisms that combine to 'abstract away'-- and thus obfuscate--such dependencies, that we are likely to encounter in modern, or recent legacy applications. To take a small example, in order to make an application localized (i.e. have its textual output adjusted to the \emph{locale} of the operating system running the application), we have to:
\begin{itemize}
	\item Replace literal output strings in program statements by named variables
	\item Specify language-specific values for those variables in external property files, with language specific file extensions
	\item Have the application read the locale of the underlying operating system during run-time, to figure out the appropriate file extension
	\item Have the application load the property file that has the appropriate file extension, and initialize the named variables with the proper language-specific values
\end{itemize}
This simple example illustrates some of the levels of indirection that we are likely to find in modern or recent legacy applications simply to find the value of a string constant that is used within the program.

Generally speaking, the code of modern or recent legacy applications is difficult to analyze for a variety of reasons:
\begin{itemize}
	\item They tend to be multi-language. For example, a typical JEE application will combine several languages, including: 1) Java, for both the server side and the client side (embedded within HTML, or JSP, or JSF tags), 2) Javascript, on the client side, embedded within HTML or JSP or JSF tags, 3) various property files, and 4) various configuration files (web.xml, ra.xml, etc.). For example, to figure out which Java method is called on the server side when the user presses a button on their browser, we typically need to scripts or files written in different languages, each with its own semantics
	\item They make heavy use of late/dynamic binding, wherever possible, through a combination of mechanisms, including:
	
\begin{itemize}
	\item Java reflexion/introspection, which enables a 'client program' to invoke the functionality of a class, for example, \emph{intensionally}, as opposed to \emph{literally/nominatively}, as shown in the following code excerpts
\begin{lstlisting}{java}
	...
	Customer myCustomer = new Customer(...);
	...
	Class itsClass = myCustomer.getClass();
	String methodName = "setName";
	...
	Class[] argumentTypes = {java.lang.String.class};
	...
	Method nameSetter = itsClass.getMethod(methodName,argumentTypes);
	
	Object[] arguments = {"John Smith"};
	
	nameSetter.invoke(myCustomer, arguments)
	...
}
\end{lstlisting}

In this case, it is difficult to guess that the method \texttt{void setName(String s)} of class \texttt{Customer} is called by this code fragment since the method is not invoked by name as in \texttt{myCustomer.setName("John Smith")}. Practically, we need to find the \emph{value} of the variable \texttt{methodName} to find which method is ultimately called\footnote{In this example, the variable \texttt{methodName} is initialized in the same scope, and the value is easy enough to find, but the value could have been returned by a function, in which case, a full dataflow analysis is needed}.
\item Data-driven control. For example, the \emph{observer} pattern introduces a level of indirection which makes it harder to guess which \emph{observer} method is called when a specific \emph{change} happens to the \emph{observable}\footnote{Recall that methods in the \emph{observable} that make changes that are worth notifying the \emph{observer} about, need to notify the observer \emph{before exiting} by calling a generic notification method as in \texttt{this.changed(changeType);} where \texttt{changeType} is a string or a value from an enumeration that is agreed upon between the \emph{observable} and the \emph{observer} that would tell the observer what to do. In one variant, the observer use a \texttt{switch} statement on the \texttt{changeType} to select the appropriate handle of change.}. Again, data flow analysis would be needed in this case to figure out the \emph{values} of the data control variables.
\item Runtime input. Since JEE applications tend to be interactive, some of the routing control will depend on user input, and thus, we won't know until run-time which path the control flow will take at a given point in the execution
\end{itemize}
\item Reliance on frameworks and container services. Such applications are typically run within \emph{containers}, which provide a number of \emph{services} (persistence, security, transaction) which imply, among other things, that specific methods from user code (so-called \emph{callback methods} will be called by the container/server, at specific points in the application lifecycle. Such call relationships will not be visible in the user code. We need to find a way of closing the gap.
\end{itemize}

\begin{figure}[h]
\centering
\includegraphics[width= 0.9\linewidth]{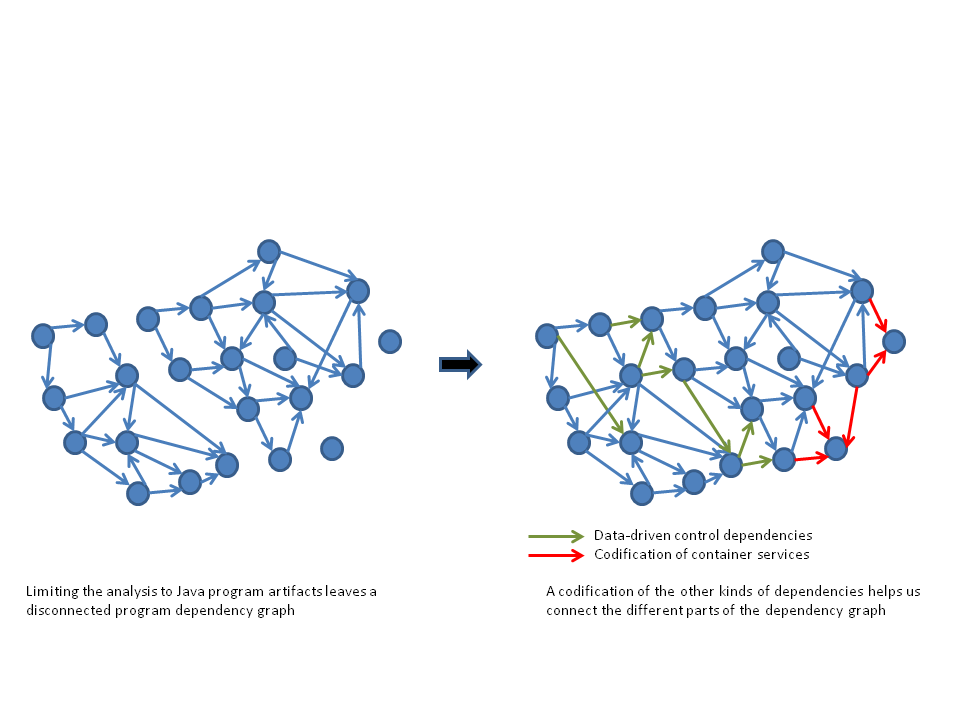}
\caption{Java code analyses have to be complemented with the analysis of, 1) other program artefacts, and 2) services offered by the run-time infrastructure}
\label{fig.encoding-different-kinds-of-dependencies}
\end{figure}

All of these considerations mean that to get a complete representation of relationships between program elements, we need to augment the traditional \emph{unilingual} program static analysis techniques with other kinds of analyses, involving other kinds of artifacts, but also, possibly, involving the codification of services offered by containers/application servers. Figure \ref{fig.encoding-different-kinds-of-dependencies} illustrates this.

In the remainder of this section, we will discuss some of these issues in more detail.

\subsubsection{Characterizing and codifying invocation patterns}
\label{characterizing-codifying-invocation-patterns}

\begin{figure}[h]
\centering
\includegraphics[width= 0.9\linewidth]{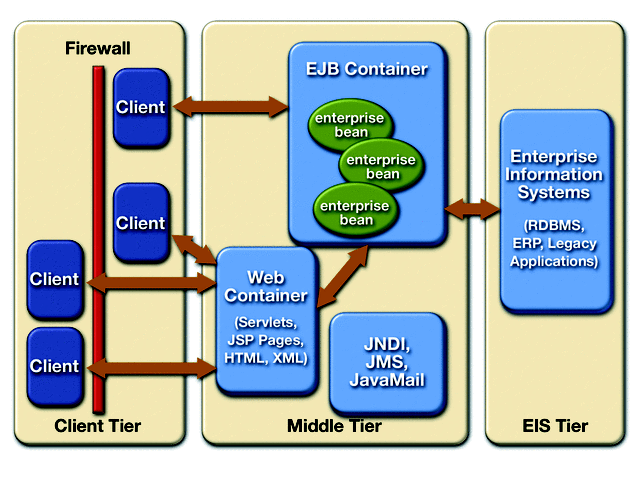}
\caption{An overview of the components of the J2EE family of technologies, circa 2002 (EJB2.x)}
\label{fig.j2ee-overview}
\end{figure}

Figure \ref{fig.j2ee-overview} shows an overview of the J2EE technology, circa 2002 (EJB 2.x). The figure shows a number of \emph{technologies} (frameworks, tools, API) that are built on top of basic Java APIs, that abstract some of the services offered by the J2EE technology. Ultimately, in a web application, when all is said and done, when the user interacts with any widget on their client interface, some (typically server-side) user java code (as opposed to infrastructure code) will be called. The different technologies that are provided, such as servelets, Java Server Pages, or Java Server Faces, provide abstraction layers enabling client-side developers to focus on the visual aspects, and wrote as little application code as possible, to link up with the server side logic.

Thus, we need to examine the different technologies and figure out how that invocation works. In \cite{Shatnawi2016} \cite{shatnawi2018implement}, we examined, among other things, invocation patterns for JSP, and identified \emph{half a dozen ways} that a JSP page can be invoked\footnote{JSP pages are mapped, by JSP containers, into \emph{servelets}. The mapping can be done at deployment time or even during runtime. We will refer interchangeably/abusively to JSP pages, and the corresponding servelets}, including:

\begin{itemize}
\item Through the web.xml file, which shows the correspondence between JSP page names, URL's, and servlet classes
\item Through annotations within Java classes (since Java 3)
\item Through different kinds of explicit calls between servelets, can be found in servelet code, or in Java code fragments embedded within JSP tags within JSP pages, or Java code fragments embedded within HTML tags within HTML pages
\item Etc. (see \cite{Shatnawi2016}).
\end{itemize}

When analyzing a JEE application, we need to look for the different invocation patterns that can be found in the different kinds of artifacts.

Similar analyses need to be made for other JEE technologies, including Java Server Faces (see \cite{Shatnawi2016} and \cite{shatnawi2018implement}). For the purpose of this project,we will not perform an \emph{exhaustive} analysis of \emph{all} the technologies. We will limit ourselves to the most commonly used technologies in the early JEE applications ('recent legacy'), and/or to the technologies needed for our experimentation.

\subsubsection{Don't call us we will call you}
\label{dealing-with-frameworks}

Unlike code libraries, which developers can (re)use by \emph{explicitly} calling the functionality that they need, \emph{applications frameworks} embody reusable functionality through a combination of, 1) services that are supplied by the framework to user code, \emph{provided that the user code implements some predefined functionality}, and 2) reusable artifacts that developers can reuse, in a way similar to libraries. The 'service contract' between the framework developer and the framework user relies on \emph{inversion of control}, or, as early OO pioneers called it, 'the Hollywood principle: don't call us we will call you'.

\begin{figure}[h]
\centering
\includegraphics[width= 0.9\linewidth]{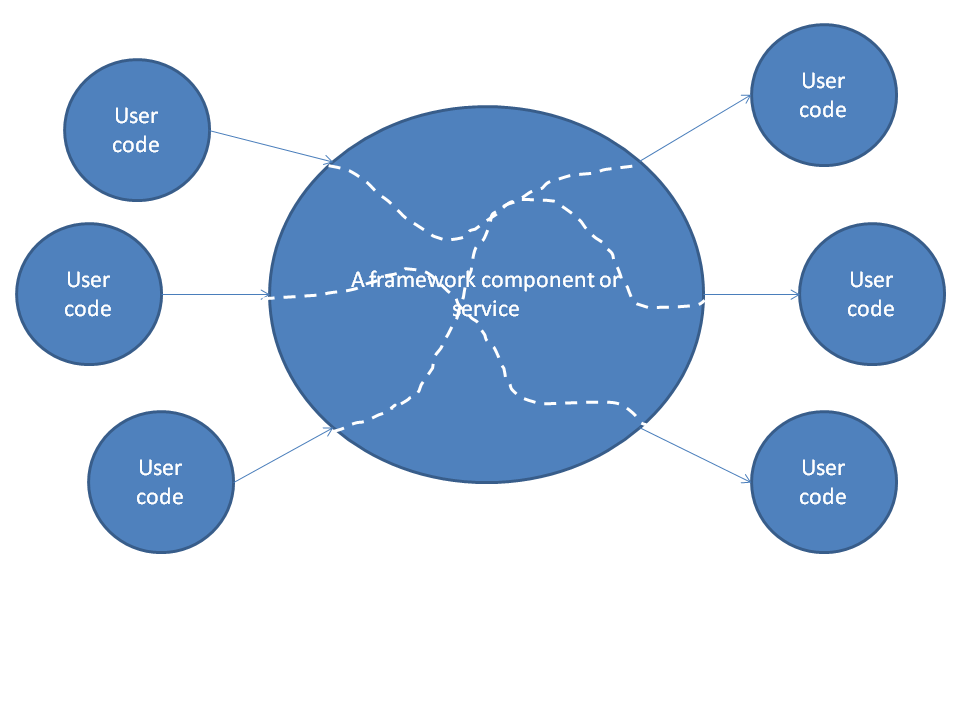}
\caption{We need to codify the dependencies that are inherent in frameworks}
\label{fig.dealing-with-frameworks}
\end{figure}

Web applications, be they JEE or otherwise, rely \emph{heavily} on frameworks to ease the development, and manage the dependencies between the various layers of the application, including MVC frameworks, and persistence frameworks. To take advantage of the functionality provided by such frameworks (e.g. linking up and synchronizing models to views and controllers), developers have to implement call back methods which will be called by the 'framework infrastructure' at specific application execution times/events. However, if we want to figure out which server-side method, say, is called when the user presses a button, I need to be able to figure out how a particular framework associates application events with user code. Thus, for a given JEE application, we need to explicitly codify/representy the dependencies that are managed by the frameworks it uses. Figure \ref{fig.dealing-with-frameworks} illustrates the idea.

Note that having the source code of the framework available does not necessarily make the problem any easier. To be able to offer a set of services to 'arbitrary' user code, frameworks tend to use, themselves, to have all of the characteristics of JEE applications as a whole: 1) multi-lingual, 2) heavy use of late binding techniques, such as the ones mentioned in Section \ref{overview-issues-static-analysis}, and the use of code generators. Not to mention that the source code is not always available. Thus, we are better off codifying the 'framework contracts' \emph{explicitly}, as opposed to detecting them through code analysis. More on this when we talk about container services.

\subsubsection{Codifying container service contracts}
\label{container-service-contracts}

JEE application servers offer a number of \emph{services} to contained/hosted applications, that \emph{do not require} end user programming, including remote method invocation, lifecycle management, persistence, security, persistence, and transactions. In fact, this was one of the selling points of the JEE technology, as a technology for enterprise-class, distributed applications, as compared to the CORBA standard, for example, which defined many of the same services, \emph{in the form of APIs that enterprise application developers needed to invoke}. Thus, it was argued, let application developers focus on the business logic, while we (JEE application servers) worry about the infrastructure services needed to support your applications.

\begin{figure}[h]
\centering
\includegraphics[width= 0.9\linewidth]{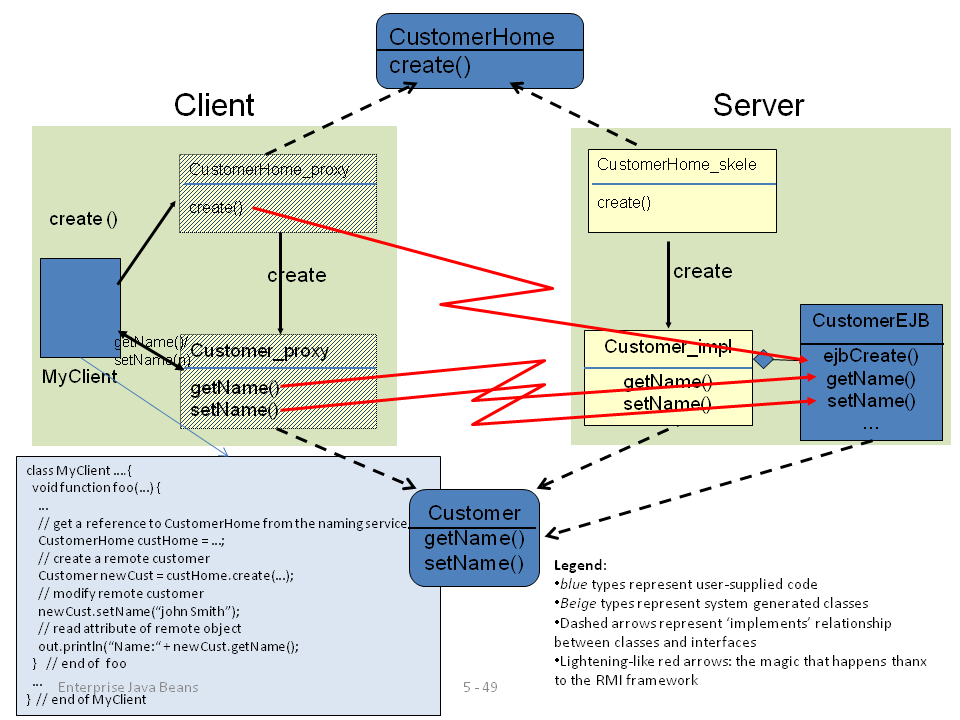}
\caption{We need to codify the dependencies that are inherent in RMI}
\label{fig.hidden-dependencies-rmi}
\end{figure}

A number of these services come out-of-the-box, and are offered to all the hosted applications, including remote method invocation and lifecycle management. Let us start with remote method invocation. Figure \ref{fig.hidden-dependencies-rmi} shows the example of a two-tier client-server JEE application that uses EJBs-- an \emph{entity bean} in this case. We assume in this case that we are developing a client application (class \texttt{MyClient}) that manipulates customer objects which are managed by a remote server. The blue boxes represent developer-supplied code, which includes, 1) the client application (\texttt{MyClient} class), and 2) the customer EJB. In turn, to develop an EJB to represent \emph{shared, remote instances} of customer, we need to supply, 1) a Java interface representing the remotely available public methods of customers (\texttt{Customer}), 2) a Java interface to represent some sort of a remote 'customer factory' (\texttt{CustomerHome}), and 3) the actual server-side class that implements customers (\texttt{CustomerEJB}).

Typically, such an application would be coded in (at least) two separate projects, one for the client side, and one for the server side. To compile the client side project, all I need is the Java interfaces (\texttt{Customer} and \texttt{CustomerHome}), and all I know is that the method \texttt{MyClient.foo(...)} is able to create an object \emph{of a class that implements} \texttt{Customer}, but I don't know which class that is.
\footnote{This is a simplification of reality, which is actually more complicated. In J2EE and J3EE, when we \emph{deploy} an EJB, the deployment tool generates, among other things, a JAR file to be included in client projects. Such a JAR file will contain the Java interfaces \emph{and the client side proxies}, represented in Figure \ref{fig.hidden-dependencies-rmi} by the shaded Java classes \texttt{Customer\_Proxy} and \texttt{CustomerHome\_Proxy}. 
If I decided to complement my analysis of the client application with those two proxies, I might be able to figure out that the variable \texttt{newCust} of method \texttt{foo(...)} is \emph{necessarily} of type \texttt{Customer\_Proxy} (for example, by looking at which class is instantiated in the \texttt{create(...)} method of \texttt{CustomerHome\_Proxy}), I would have no way of finding that the method \texttt{getName()} of \texttt{Customer\_Proxy} calls the method \texttt{getName()} of the server side class \texttt{CustomerEJB}. The reason for that is that prior to 'traveling down the wire', method calls are packed into generic request invocation objects (that include identity of remote object, method name, and method parameters), which are \emph{unpacked at the other/server end by the ORB (Object Request Broker)}, which \emph{locates the appropriate instance of} \texttt{Customer\_Impl}, which then delegates the call the \texttt{CustomerEJB}.}
 
Similarly, on the server side, I may have a bunch of classes and methods that seem to be waiting to be instantiated and/or called, but I have no code to instantiate those classes/call those methods. The missing link \emph{can't even be identified if I had the source code of the server}, because the server makes heavy use of the Java reflection API (see footnote). Thus, the only way to identify the missing links is \emph{by understanding the underlying technology and by codifying its services explicitly}. In this particular case, having understood how remote method invocation works in JEE, I can \emph{safely and surely} add a call link between the method \texttt{MyClient.foo(...)} and the method \texttt{CustomerEJB.setName(...)}. In fact, I don't even need to worry about polymorphism or the runtime type of (the server side) \texttt{newCust}: it can \emph{only be} an instance \texttt{CustomerEJB}. This example illustrates the kind of work that we need to do for JEE services.

\begin{figure}[h]
\centering
\includegraphics[width= 0.9\linewidth]{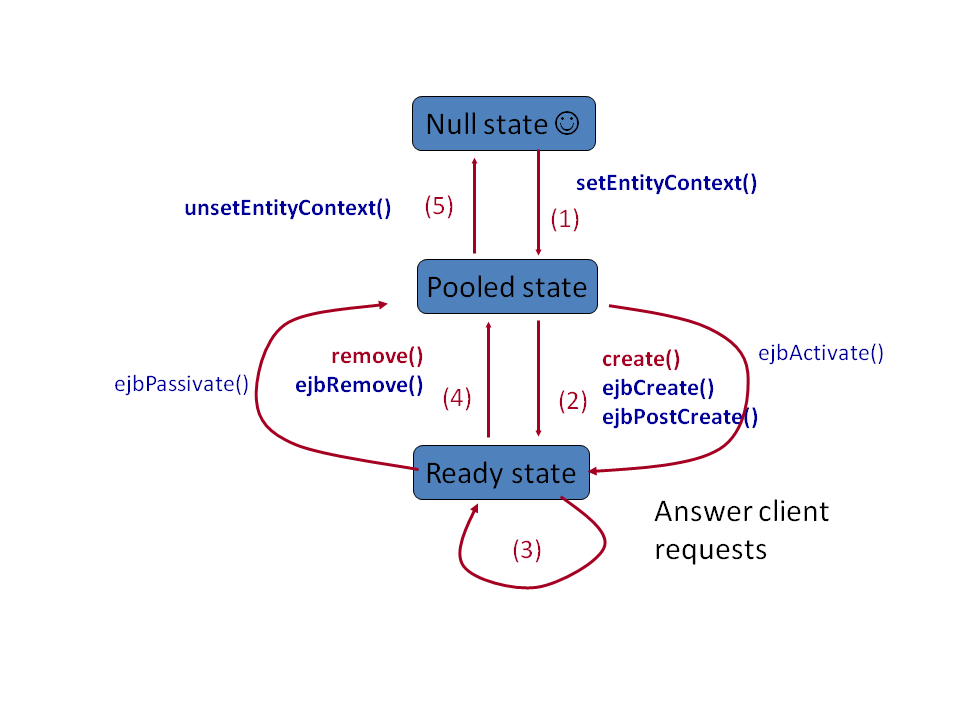}
\caption{The lifecycle of EJBs. Red methods belong to the \emph{home} interface, but they end up invoking the blue methods on the EJB class (e.g. a call to create() on the home object lead to a call to ejbCreate() and ejbPostCreate() on the EJB object). The bold methods correspond to method invoked following explicit user actions. Non-bold methods are invoked by the server to manage its resources}
\label{fig.ejb-lifecycle}
\end{figure}

Another example of JEE container services is the \emph{lifecycle management} service, which will manage the lifecycle of EJBs in a way that optimizes resource usage within the server. Figure \ref{fig.ejb-lifecycle} shows the lifecycle of \emph{entity} beans, and it provides examples of the funny things servers do. First, we have to distinguish between three states for an entity bean: 1) null-state, which is the initial and final state, corresponding to the instance 'not-existing', 2) the \emph{pooled state}, where an instance is created and is made available to service client applications, and 3) the \emph{ready state}, where the object is assigned to a specific business object, and is ready to service client application requests. To understand the meaning of \emph{pooled state}, we should mention that JEE servers maintain \emph{pools of EJB objects} during runtime, that they may assign to specific client programs. This is done, in part to save on object instantiation time, which is typically time-consuming, especially for large objects. Thus, when an object is no longer needed (for example, the shopping cart from your amazon session), instead of garbage-collecting it, the server "empties its fields", and returns it to the pool: that is the purpose of the \texttt{remove()} method. It is similar to the Java \texttt{release()} method, in the sense that you are supposed to reset the field of the object (to remove hanging references), and to release any resources it may hold (database connections, files, etc.). The \texttt{remove()} method on the home object actually calls the \texttt{ejbRemove()} on the EJB object. Similarly, when a client application ask a 'home object' to create an entity bean, the server pulls one from the pool, initializes its fields based on the parameters of \texttt{create(...)} (for example, a customer ID). That call will provoke a call to \texttt{ejbCreate(...)} with the same parameters (the same customer ID), which, in \emph{bean managed persistence}, can be used to query the database and load the other fields. The method \texttt{ejbPostCreate()} acquires any additional resources that are needed, beyond object loading/initialization. It is at this point that a \texttt{CustomerEJB} object is 'loaded' with the data for an actual customer from the database, and is able to respond to queries.

When the server runs low on memory, it will call the \texttt{ejbPassivate(...)} method to typically \emph{serialize} the EJB object on hard disk, so that it may be activated later. Serialization and deserialization to/from files (e.g. XML) are typically faster than executing a SQL query, and loading the object from the result set, especially because we can serialize and deserialize computed attributes, which are \emph{not} stored in the database. This can lead to interesting design trade-offs between memory and CPU usage, as we weigh how much to save/serialize during passivation, against how much we need (and can) recompute during activation. But it illustrates a 'difficulty' with containers:
\begin{itemize}
	\item They will call some methods without 'telling us' when. We cannot precisely characterize 'when the server runs low on memory'.
	\item Some of these methods may contain (or invoke) quite a bit of business logic. For example, if we decide to not serialize computed/recoverable attributes during \texttt{ejbPassivate(...)}, we need to recompute those attributes in \texttt{ejbActivate()}.
\end{itemize}

Generally speaking, container services fall into two general categories:
\begin{itemize}
	\item out of the box services that apply to all deployed applications, including RMI and lifecycle management
	\item configurable services such as persistence, transactions, and security, where the parameters of the service can be found in many places
\end{itemize}

For example, with transaction services, the developer can specify which of the methods of an EJB need to be executed within transaction boundaries. This can be done in one of two ways: 1) in the EJB deployment descriptors ('old style'), or 2) as code annotations of the 'transaction methods'. Similarly, the \emph{security service} enables developers to specify restrictions on the credentials of the client applications that call a particular method. Such restrictions can be specified in a security policy file, for coarse-grained security policies, or in deployment descriptors.

We make an additional distinction between container services:
\begin{itemize}
	\item Services that are configured/specified \emph{extensionally}, i.e. by listing the \emph{explicitly} the entities/software components that benefit from the service. The security and transaction service are examples of such \emph{extensionally specified} services, and
	\item Services that are specified \emph{intensionally}, for example in terms of properties that the hosted applications/software components must satisfy to take advantage of the service. For example, all classes that implement a particular interface, or that extend a particular class, will take advantage of the service. Examples of such services are RMI or lifecycle management. \footnote{For example, the lifecycle of an EJB depends on whether it is an \emph{entity bean} or a \emph{session bean}. The nature of the EJB is specified in the deployment descriptor, but is also implicit in the interfaces/classes that are implemented/extended by the EJB components.}
\end{itemize}

The difference between the last two categories is important, for our purposes, because we will need to look in two different places to figure out whether we need to add a dependency or not. In the first case, we look at annotations and deployment descriptors. In the second case, we submit the software artifacts to a number of checks to figure out if they qualify for the service or not, and if they do, we add the dependency.

\section{Issues in Service Packaging}
\label{issues-service-packaging}
Once we have identified a cluster of functionality as a candidate service, the next step consists of packaging that cluster 'behind a service' interface. We can think of the service identification step as recognizing those clusters of functionality that are \emph{useful}, whereas service packaging as the step of making those clusters \emph{usable}. Broadly speaking, there are two sets of issues:
\begin{itemize}
	\item specifying/computing the service interfaces, based on the 'contents' of the service, and how it is used. For example, if we determine that two classes A and B, each with its set of methods, make up a potential service, we need to determine which methods of classes A and B need to be exposed as service functions.
	\item packaging them in appropriate service interfaces. Having determined which functionality needs to be exposed (previous step), now the issue is to figure out \emph{how} to expose it. This involves a number of \emph{design} and \emph{implementation}-level issues.
\end{itemize}
We will discuss the two issues in turn in the next two subsections.
\subsection{Computing service interfaces}
\label{computing-service-interfaces}
At a basic level, given a bunch of Java classes that offers a cluster of functionality that was deemed worth (re)using as a whole, we need to figure out:
\begin{itemize}
	\item the set of functions that are \emph{offered} by the cluster. These will be presented as \emph{provided interfaces}
	\item the set of functions that are \emph{needed} by the custer. These will be presented as \emph{required interfaces}
\end{itemize}
A naive approach could compute the \emph{provided interface} as the set of all the methods within the functional cluster that are called by methods that are \emph{outside} the cluster. Similarly, we could compute the \emph{required interface} as the set of all methods called by the code of the classes of the cluster, that do \emph{not} belong to the cluster. However, the resulting interfaces would not be very accurate, or useful, we explained below.

With regard to the \emph{provided interface}, there are two problems with the naive approach. First, some classes may provide functionality that is \emph{not} related to the service at hand. Including all the methods called from the outside would clutter the provided interface with methods that have little relation with the functionality provided by the service. Thus, perhaps instead of including \emph{all} of the methods of the classes that are part of the candidate service, we will only include the ones that contributed to the specific service pattern that was used to identify the candidate service. For example, if we are dealing with an interceptor-type technical service (see Section \ref{service-taxonomy}), the provided interface will be limited to those methods that have high fan-in, ignoring the others.

The second problem is more of a usability problem: a service will typically provide \emph{several} interfaces, each one of which consisting of a subset of methods, exhibiting a stronger cohesion than the whole. Thus, we have to identify 'subclusters', within the provided interface of the service, and expose those are separate provided services. This can be framed as an optimization problem, where we need to balance the size of individual interfaces (not too big), with the number of interfaces (not too big either).

\subsection{Designing and implementing service interfaces}
\label{designing-service-interfaces}
Once we have identified the functions that are part of the interfaces, we need to find a way of exposing them. Broadly speaking, we need to hide the functionality of the service behind one or several façades. A naive approach would, for example, map each method \emph{f} of class C with parameters p1,..., pi (\texttt{f(T1 p1,T2 p2,...,Ti pi)}) to a method \texttt{g(C p0, T1 p1,..., Ti pi)} of the façade, defined as follows:
\begin{lstlisting}{java}
	public class MyServiceFacade {
	
		...
		public T g(C p0,T1 p1, ..., Ti pi) {
			return p0.f(p1,...,pi);
		}
		...
	}
\end{lstlisting}
However, we need to take into account a number of issues, some related to API design, in general, while others are specific to the implementation of the façade pattern.

For example, in general API (or library) design, it is usually recommended to have functions pass simple data types, as opposed to complex data structures. This insures, among other things, portability of the functionality. For example, this was one of the design principles behind the X-Windows interfaces. To guarantee (or simplify) its portability across operating systems, its designers chose to pass the representation of various operating system objects (e.g. files or processes), as simple/individual attributes, instead of complex structures. For example, instead of passing \emph{file} objects as C structures (\texttt{struct}), the functions that manipulated files took several \texttt{int} and \texttt{char*} parameters to represent the properties of a file (e.g. name, path, file type, etc.). This made it possible to have each operating system use its own representation that is accessed internally by the code of the functions. 

If we apply the first pattern, that means that we should replace object parameters by a list of the object attributes. However, not all attributes may be needed by the specific functions. Thus, if we wanted only to pass the data that is needed by the function ... we need to know what data that is. This becomes a slicing issues: checking which slice of an object is read/written by a specific method, and pasing only the attributes of the slice as parameters.

The issue of passing the object, versus its attributes, may depend on the service type. Could it be that with \emph{business services}, we can pass the attributes--typically an object ID--where as with technical services we pass the entire object? Let us take the example of a business service for processing insurance claims, which checks the claimed expenses against the coverages included in the policy. Figure \ref{fig.claim-processing-model} shows excerpts of the relevant object model. It is a common design practice for such remote services to \emph{not} pass the entire \texttt{InsurancePolicy} object, along with the \texttt{Claim} object, but simply pass the insurance policy \emph{identifier}, and let the claim processing service fetch the policy object from persistent storage, if it needs to. This way, the policy object does not 'travel down the wire'\texttt{Large objects consume time and space to serialize, send, and deserialize at the receiving end}. The next code excerpts show what the insurance claim class might look like to make sure that insurance policy objects do not travel.
\begin{lstlisting}{java}
	public class InsurancePolicy {
		...
		private String id;
		...
	}
	
	public class InsuranceClaim {
	
		...
		private String insurancePolicyId;
		
		/**
		 * transient fields are neither serialized nor 
		 * persisted. They may be 'populated' from 
		 * persistent storage on demand/at some point, 
		 * using insurancePolicyId, and accessed in a 
		 * read mode (unless invalidated by some other 
		 * mechanism, if modified)
		 */
		private transient InsurancePolicy policy;
		...
	}
\end{lstlisting}

One could also argue that technical services \emph{typically} need the entire object, including, for example, serialization, and persistence.

In addition to the above, the fa\,cade pattern raises its own set of issues. One of the issues is the relationship between the service client programs, and the classes that are behind the service fa\,cade. To go back to our methods \texttt{f(T1,T2,...,Ti)} and \texttt{g(C,T1,T2,...,Ti)}, if we make the object of type \texttt{C} as a parameter of \texttt{g(...)}, does that mean that clients of the service now need to 'create their own' instances of \texttt{C} to pass them as parameters of \texttt{g(...)}? if so, the service fa\,cade does not sound very helpful; it is actually more cumbersome. The answer depends, in part, on whether a \emph{single} instance of \texttt{C} is needed to serve/service all of the clients of method \texttt{g(...)}, or whether each client needs its own instance of \texttt{C}:
\begin{itemize}
	\item if a single instance of \texttt{C} can serve all the clients of the service, actually we do not even \emph{need} to pass the \texttt{C} instance as a parameter to \texttt{g(...)}: we could have the façade manage a singleton, and then delegate the calls to that singleton, as shown below:
\begin{lstlisting}{java}
	public class MyServiceFacade {
	
		private static C singleton = C.getInstance();
		...
		public T g(T1 p1, ..., Ti pi) {
			return singleton.f(p1,...,pi);
		}
		...
	}
\end{lstlisting}

\item each client has its own instance of \texttt{C}. Strictly speaking, even in this case, we do not need to have the clients manage the \texttt{C} objects that serve them, or pass the corresponding object as a parameter. Clients may ask the façade to create and manage the \texttt{C} objects for them, and the façade ensures that they will be served/serviced by their assigned \texttt{C} objects. The following code excerpts illustrate this:
\begin{lstlisting}{java}
public class MyServiceFacade {
  private HashMap<Object,C> 
  		clientsServers = new HashMap<Object,C>();
  ...
  public void assignMeAC(Object client) {
	C assignedC = clientsServers.get(client);
	if (assignedC == null) 
	// the client has no C instance assigned to
	// it yet
	{
	  // I create an instance of C specifically 
	  // for the client
	  assignedC = new C(client);
	  // register it for future reference
	  clientsServers.put(client,assignedC);
	}
  }
		
  /**
   * here g(...) requires that we pass the 
   * client/caller, as a parameter
   */
  public T g(Object client, T1 p1, T2 p2, ..., Ti pi) 
  			throws NoCServerAssignedException {
	// first get the assigned C server
	C assignedC = clientsServers.get(client);
			
	// if the client has not yet been assigned 
	// a C instance, throw an exception
	if (assignedC == null) throw 
    		new NoCServerAssignedException(client);
			
	// else, execute f(...)
	return assignedC.f(p1,p2,...,pi);
  }
		
  /**
   * @deprecated
   */
  public C whoIsMyC(Object client) {
	return clientsServers.get(client);
  }
}
\end{lstlisting}
Strictly speaking, the method \texttt{whoIsMyC()} may not be needed. But if we worry about forgetting to hide a needed \texttt{C} method behind the facade interface, we can provide this method for convenience, and mark it as deprecated to discourage its use.
\end{itemize}
A somewhat contrived example of the 'one object serves all' case of a remote security manager, where one object serves all clients. An example of the one-to-one case is how a JEE container assigns a different remote \emph{session bean} to each active client.

There may be other problems related to the use of the façade pattern, since hiding classes behind façade may restrict the possibilities of their reuse. Kiczales and Lamping have shown in \cite{kiczales1992} that designers and users of class libraries may, inadvertently/abusively, rely on implementation details \emph{inherent in class/subclass relationhips} within class libraries, which tightly couples 'client programs' (class library users) to implementation details that are meant to be hidden, and that library developers are not bound to respect when they evolve the library, which may break existing code. A similar situation may occur with façades.

\section{Issues in Refactoring}
\label{issues-refactoring}
Now that we have identified functional clusters as candidate services (see Section \ref{issues-service-identification}), and that we have wrapped that functionality behind services interfaces (see Section \ref{issues-service-packaging}), we need to refactor existing applications that \emph{use} that cluster of functionality, so that they invoke the functionality through the newly lstlisting service interfaces. This raises a number of issues, which we will touch upon below.

The first issue has to do with the scope of the refactoring. Assume that the method \texttt{T f(T1 p1, ...,Ti pi)} of class \texttt{C} has been identified as part of a service \texttt{S} that was packaged under a façade, as explained in Section \ref{designing-service-interfaces}. Should I consider \emph{all} invocations of \texttt{f(...)} as being part of a service call, and reroute that call through the service interface? That is a good question. For one thing, the class \texttt{C} could be part of several services, or could be used both within and outside the service. If we hide \texttt{C} behind the service interface, could clients manage? if we decide \emph{not} to include all of the methods of \texttt{C} in the service interface of \texttt{S}, then a first-cut decision could be to route calls to service-included methods to the service, and leave the others untouched. Even so, there may be (definitely \emph{is}) an overhead in calling a service interface, which could be remote, as opposed to manipulating the \texttt{C} class directly. If the class \texttt{C} is part of some library, then we are in trouble. For example, a business service that computes the yearly repayment of a loan of based on some capital amount, a yearly interest rate, and an amortization period, might expose a 'utility' method that computes the compound interest rate over a number of periods. Should have every call to that method go through the service interface, or should I make it possible to invoke the method independently of the service?

Part of the answer \emph{may lie} in the invocation pattern of the caller: what other methods of the service, if any, does it call, in addition to method \texttt{f(...)}. Is there a typical invocation sequence that uniquely characterizes the functionality of the service? For example, a flight booking service may offer methods to, 1) search for flights (call it \texttt{searchFlights(...)}, 2) to get the details of a flight (\texttt{getFlightDetails(...)}, and 3) to book a seat at a particular flight (\texttt{bookFlight(....)}). I can imagine the \texttt{searchFlights(...)} method as being part of an aggregation service that does price comparison. When I find a call to \texttt{searchFlights(...)}, should I consider it as an invocation of the booking service, or of the comparative shopping service. Part of the answer may lie in which other methods were called before or after. Thus, to be able to characterize a method call as being part of the invocation of a service, I may need to check whether other functions of the service (the façade) are invoked, within a certain scope from the first call. Dynamic traces would make this pattern detection easier--provided that we can characterize the typical, or expected, or permissible invocation sequences.

Once I have determined that a client call to a method that was included in a service is a call to that service, I need to focus on the mechanics of the code transformation itself. Depending on what \emph{flavor} of service-orientation we use, but the refactoring may involve a \emph{paradigm shift}. For example, in going from Java EE to SOA, I might chose an event-based implementation of SOA, in which case I needed to migrate some functional code from a \emph{call-and-return} style, to an asynchronous \emph{event-based style}. This presents its own challenges:
\begin{itemize}
	\item Going from structural/flow-based functional composition to data-level composition:
\begin{itemize}
	\item with structural/flow based composition, function calls are hardcoded in the program--reflection notwithstanding
	\item with data-level composition, composition happens through shared messages.
\end{itemize}
Anyone who has coded variants of the \emph{observer} pattern knows the challenges of implementing it in such a way as to communicate \emph{complex} changes that may involve \emph{complex} data structures/objects, and doing so through a simple interface\footnote{Coding the \texttt{changed(...)} method of the \texttt{Observable} requires some ingenuity to communicate complex changes and complex data through two parameters, as in \texttt{this.changed(changeAspect,changeData)}. A realistic implementation of the \texttt{Observer} either requires different variants of the \texttt{changed(...)} method, or a fairly complex representation of the change data, which tightly couples the \texttt{Observer} which needs to know how to unpack the change data for specific change aspects, defeating the purpose of the \emph{observer} pattern}.
\item Going from a synchronous call style to an asynchronous call style. This involves two challenges, 1) at the algorithmic level, 2) the architectural level.
\end{itemize}
At the algorithmic level, we illustrate the kind of changes that we will need to make. Consider the function \texttt{f(...)} that invokes functions \texttt{g(...), h(...), i(...)}, which are now part of service interfaces. We will ignore object-orientation because the issues are the same, whether we use procedural code or object-oriented code.

\begin{lstlisting}{java}
public T1 f(T2 x) {
	T3 y = g(x);
	T4 z = h(y);
	T1 u = i(z);
	return u;
}
\end{lstlisting}

Here, we have a simple composition where the output produced by \texttt{g(.)} is used as an input to \texttt{h(.)}, whose output is used as input to \texttt{i(.)}, whose output is returned by \texttt{f(.)}. Thus \texttt{f(x) = i(h(g(x)))}.
If we were to transform \texttt{f(.)} into a message driven style (e.g. \emph{à la} Java Messaging Service), we would have each of \texttt{g(.)}, \texttt{h(.)}, and \texttt{i(.)} receive their inputs through named input queues, and put their outputs on named output queues. The handle the coordination between them, we would then need an \emph{message/event listener}, then listens for messages arriving on queues, and takes the appropriate action. Thus, the code for performing this composition would look something like the following:

\begin{lstlisting}{java}
public void onMessage(aMessage) {
	// first get the message type to figure
	// out where to send it next
	int messageType = aMessage.getType();
	
	// then switch on the message type
	switch (messageType) {
		case F_REQUEST:
			// unpack aMessage and get x value
								
			x = aMessage.get();
									
			// create an G message to send x to G next
			anG_Message = createG_Message(x);
									
			// put anG_Message on G in_queue;
			...
			return;
		case G_RESPONSE: 
			// unpack aMessage and get y value
									
			y = aMessage.get();
									
			// create an H message to send y to H next
			anH_Message = createH_Message(y);
									
			// put anH_Message on H in_queue;
			...
			return;

		case H_RESPONSE:
			// unpack aMessage to get the z value
			...
			z = aMessage.get(...)
									
			// create an I message to send z to I next
			anI_Message = createI_Message(z);
									
			// put anI_Message on I in_queue
			...
			return
}\end{lstlisting}

This example illustrates how cumbersome it is to implement a simple sequence in the JMS style. Typically, the functions that are accessible through message queues tend to be of relatively coarse granularity. From an architectural point of view, if you use an asynchronous/message-driven style, at a particular level of functional aggregation, it pretty much imposes that style at the \emph{levels above it}. In this case, it means that if \texttt{f(...)} is implemented in a message-oriented way, then its callers will \emph{likely} need to be, themselves, message oriented\footnote{If the caller of \texttt{f(...)} does not care about the output produced by it, but just needs to make sure that it is called, then it need not be asynchronous/message oriented. I can have a blocking function call launch an independent thread and return, considering its job is done. That thread (\texttt{f(...)} in this case) will eventually terminate and do whatever needed to be done. But if the caller needs the output of \texttt{f(...)} to continue, then it \emph{too} needs to be message-oriented}.

Of course, you are not going to code a sorting algorithm this way. But this raises interesting issues in terms of execution sequence/trace equivalence, and error handling, among other things.
\section{Conclusion}
\label{conclusion}

In this paper, we discussed issues and the state of the research related to the re-engineering of legacy JEE applications into service oriented ones. We argued that this migration requires that we perform three steps:
\begin{enumerate}
\item Identify clusters of functionality within the legacy application(s) that qualify as candidate services
\item Package these functional clusters into a service-like packaging so that enterprise applications can invoke these functionalities using the service interfaces, and
\item Refactor existing applications so that they invoke these 'newly lstlisting' services using their service interfaces, thereby increasing their future maintainability.
\end{enumerate}
It is fair to say that these steps received attention in the literature by decreasing order. 

A lot of work has been performed on component and service identification. But like we mentioned in section 3, most of the existing work looks for services based on their functional cohesion and low coupling with other parts of business applications, regardless of service types. We argued that different service types have different code smells/signatures in legacy code, and hence, service identification methods should take into account service types. Thus, we proposed a service taxonomy in section \ref{service-taxonomy}, and proposed a preliminary set of service-type specific identification criteria in section \ref{service-identification-approach}. As our approach relies on source code, it becomes important to identify all of the static dependencies within legacy applications. We identified a number of issues related to the static code analysis of JEE applications, including the fact that they are multi-tiers, multi-language, their reliance on a number of 'extra-lingual' mechanisms for linkage and bindings, and their reliance on frameworks and containers that hide some call dependencies. Thus, prior to looking for functional clusters that can qualify as candidate services, we need to map out all the dependencies that are inherent in a JEE application, beyond ones that static analysis of Java code can find.

With regard to service packaging, discussed in section \ref{issues-service-packaging}, we identified a number of issues that need to be resolved, which can think of as \textit{interface specification} (section \ref{computing-service-interfaces}, and \textit{interface design} (see section \ref{designing-service-interfaces}). Roughly speaking, specification deals with ways to partition the set of functions offered or consumed by a functional cluster into a set of relatively independent and more strongly cohesive interfaces so that client programs may be exposed to--and use--those interfaces, as opposed to the full gamut of functions available. We discussed some potential approaches to do this. Interface \textit{design} deals with the detailed design of individual service function \textit{signatures}, and encompasses things such as granularity of function parameters, and lifecycle management of objects that live within the service.

With regard to the refactoring of existing applications, we barely started exploring the relevant issues. Basically, there are two sets of issues: 
\begin{enumerate}
\item The client side consequence of service interface design. Indeed, as we hide functional parameters, service-side object creation and lifecycle management, client programs need to manipulate the service functionalities through the newly defined (and constraining) service interfaces
\item The general issue of changing architectural style in the migration process. In particular, we looked at the specific example of going from a call-and-return invocation style--prevalent in JEE applications--to a message-oriented style\footnote{Technically speaking, JMS and Message-Driven Beans are also part of JEE}, to identify some of the relevant issues.
\end{enumerate}
In this report, we raised more questions than we answered, and that was the intent. Our priority is on identifying the problems. In those cases where our thinking has progressed, we presented the first elements of an approach, but for the most part, all of the research problems are still fairly open.

More reports--and publications--will follow this report, by expanding on some of the issues, and proposing a solution approach, or elaborating on some of the elements presented here. We will not evolve this report: progress that we make on any topic will be described in separate reports. 

\section*{References}
\bibliographystyle{Style}
\bibliography{main.bib}

\begin{thebibliography}{21}
\expandafter\ifx\csname natexlab\endcsname\relax\def\natexlab#1{#1}\fi
\providecommand{\bibinfo}[2]{#2}
\ifx\xfnm\relax \def\xfnm[#1]{\unskip,\space#1}\fi
\bibitem[{Suwisuthikasem and Samadzadeh(2015)}]{Suwisuthikasem2015}
\bibinfo{author}{S.~Suwisuthikasem}, \bibinfo{author}{M.~H. Samadzadeh},
  \bibinfo{title}{Migration from Legacy Systems to SOA Applications: A Survey
  and an Evaluation}, \bibinfo{publisher}{Springer International Publishing},
  \bibinfo{address}{Cham}, pp. \bibinfo{pages}{609--614}.
\bibitem[{Cai et~al.(2011)Cai, Liu, and Wang}]{Cai2011}
\bibinfo{author}{S.~Cai}, \bibinfo{author}{Y.~Liu}, \bibinfo{author}{X.~Wang},
\newblock \bibinfo{title}{A survey of service identification strategies},
\newblock in: \bibinfo{booktitle}{IEEE Asia-Pacific Services Computing
  Conference}, pp. \bibinfo{pages}{464--470}.
\bibitem[{Mili et~al.(1995)Mili, Mili, and Mili}]{Mili1995}
\bibinfo{author}{H.~Mili}, \bibinfo{author}{F.~Mili},
  \bibinfo{author}{A.~Mili},
\newblock \bibinfo{title}{Reusing software: Issues and research directions},
\newblock \bibinfo{journal}{IEEE Trans. Softw. Eng.} \bibinfo{volume}{21}
  (\bibinfo{year}{1995}) \bibinfo{pages}{528--562}.
\bibitem[{Mili et~al.(2001)Mili, Mili, Yacoub, and Addy}]{Mili2001}
\bibinfo{author}{H.~Mili}, \bibinfo{author}{A.~Mili},
  \bibinfo{author}{S.~Yacoub}, \bibinfo{author}{E.~Addy},
  \bibinfo{title}{Reuse-based Software Engineering: Techniques, Organization,
  and Controls}, \bibinfo{publisher}{Wiley-Interscience}, \bibinfo{address}{New
  York, NY, USA}, \bibinfo{year}{2001}.
\bibitem[{Erl(2007)}]{Erl2007}
\bibinfo{author}{T.~Erl}, \bibinfo{title}{SOA Principles of Service Design},
  \bibinfo{publisher}{Prentice Hall PTR}, \bibinfo{address}{Upper Saddle River,
  NJ, USA}, \bibinfo{year}{2007}.
\bibitem[{Cohen(2007)}]{Shy2007}
\bibinfo{author}{S.~Cohen},
\newblock \bibinfo{title}{Ontology and taxonomy of services in a
  service-oriented architecture},
\newblock \bibinfo{journal}{The Architecture Journal} \bibinfo{volume}{11}
  (\bibinfo{year}{2007}) \bibinfo{pages}{30--35}.
\bibitem[{Ope(????)}]{OpenGroup}
\bibinfo{title}{The open group soa reference architecture},
\newblock in:
  \bibinfo{booktitle}{\url{http://www.opengroup.org/soa/source-book/soa_refarch/}}.
  \bibinfo{note}{[Online; accessed 19-July-2017]}.
\bibitem[{Fuhr et~al.(2011)Fuhr, Horn, and Riediger}]{Fuhr2011}
\bibinfo{author}{A.~Fuhr}, \bibinfo{author}{T.~Horn},
  \bibinfo{author}{V.~Riediger},
\newblock \bibinfo{title}{Using dynamic analysis and clustering for
  implementing services by reusing legacy code},
\newblock in: \bibinfo{booktitle}{18th Working Conference on Reverse
  Engineering}, pp. \bibinfo{pages}{275--279}.
\bibitem[{Alahmari et~al.(2010)Alahmari, Zaluska, and De~Roure}]{Alahmari2010}
\bibinfo{author}{S.~Alahmari}, \bibinfo{author}{E.~Zaluska},
  \bibinfo{author}{D.~De~Roure},
\newblock \bibinfo{title}{A service identification framework for legacy system
  migration into soa},
\newblock in: \bibinfo{booktitle}{2010 IEEE International Conference on
  Services Computing}, pp. \bibinfo{pages}{614--617}.
\bibitem[{Zhang et~al.(2005)Zhang, Liu, and Yang}]{Zhang2005e}
\bibinfo{author}{Z.~Zhang}, \bibinfo{author}{R.~Liu},
  \bibinfo{author}{H.~Yang},
\newblock \bibinfo{title}{Service identification and packaging in service
  oriented reengineering},
\newblock in: \bibinfo{booktitle}{SEKE}, volume~\bibinfo{volume}{5}, pp.
  \bibinfo{pages}{620--625}.
\bibitem[{Grieger et~al.(2014)Grieger, Sauer, and Klenke}]{Grieger2014}
\bibinfo{author}{M.~Grieger}, \bibinfo{author}{S.~Sauer},
  \bibinfo{author}{M.~Klenke},
\newblock \bibinfo{title}{Architectural restructuring by semi-automatic
  clustering to facilitate migration towards a service-oriented architecture},
\newblock \bibinfo{journal}{Softwaretechnik-Trends} \bibinfo{volume}{34}
  (\bibinfo{year}{2014}).
\bibitem[{Gu and Lago(2010)}]{Gu2010}
\bibinfo{author}{Q.~Gu}, \bibinfo{author}{P.~Lago},
\newblock \bibinfo{title}{Service identification methods: A systematic
  literature review},
\newblock in: \bibinfo{booktitle}{Third European Conference ServiceWave
  Proceedings}, \bibinfo{publisher}{Springer Berlin Heidelberg},
  \bibinfo{year}{2010}, pp. \bibinfo{pages}{37--50}.
\bibitem[{Huergo et~al.(2014)Huergo, Pires, Delicato, Costa, Cavalcante, and
  Batista}]{Huergo2014}
\bibinfo{author}{R.~S. Huergo}, \bibinfo{author}{P.~F. Pires},
  \bibinfo{author}{F.~C. Delicato}, \bibinfo{author}{B.~Costa},
  \bibinfo{author}{E.~Cavalcante}, \bibinfo{author}{T.~Batista},
\newblock \bibinfo{title}{A systematic survey of service identification
  methods},
\newblock \bibinfo{journal}{Serv. Oriented Comput. Appl.} \bibinfo{volume}{8}
  (\bibinfo{year}{2014}) \bibinfo{pages}{199--219}.
\bibitem[{Tonella and Ceccato(2004)}]{Tonella2004}
\bibinfo{author}{P.~Tonella}, \bibinfo{author}{M.~Ceccato},
\newblock \bibinfo{title}{{Aspect mining through the formal concept analysis of
  execution traces}},
\newblock in: \bibinfo{booktitle}{Proceedings - Working Conference on Reverse
  Engineering, WCRE}.
\bibitem[{Robillard and Murphy(2002)}]{Robillard2002}
\bibinfo{author}{M.~Robillard}, \bibinfo{author}{G.~Murphy},
\newblock \bibinfo{title}{{Concern graphs: finding and describing concerns
  using structural program dependencies}},
\newblock \bibinfo{journal}{Proceedings of the 24th International Conference on
  Software Engineering. ICSE 2002}  (\bibinfo{year}{2002})
  \bibinfo{pages}{406--416}.
\bibitem[{Marin et~al.(2006)Marin, Moonen, and {Van Deursen}}]{Marin2006}
\bibinfo{author}{M.~Marin}, \bibinfo{author}{L.~Moonen},
  \bibinfo{author}{A.~{Van Deursen}},
\newblock \bibinfo{title}{{A common framework for aspect mining based on
  crosscutting concern sorts}},
\newblock in: \bibinfo{booktitle}{Proceedings - Working Conference on Reverse
  Engineering, WCRE}, pp. \bibinfo{pages}{29--38}.
\bibitem[{{El Boussaidi} et~al.(2012){El Boussaidi}, Belle, Vaucher, and
  Mili}]{ElBoussaidi2012}
\bibinfo{author}{G.~{El Boussaidi}}, \bibinfo{author}{A.~B. Belle},
  \bibinfo{author}{S.~Vaucher}, \bibinfo{author}{H.~Mili},
\newblock \bibinfo{title}{{Reconstructing architectural views from legacy
  systems}},
\newblock in: \bibinfo{booktitle}{Proceedings - Working Conference on Reverse
  Engineering, WCRE}, pp. \bibinfo{pages}{345--354}.
\bibitem[{Belle et~al.(2014)Belle, {El Boussaidi}, and Mili}]{Belle2014}
\bibinfo{author}{A.~Belle}, \bibinfo{author}{G.~{El Boussaidi}},
  \bibinfo{author}{H.~Mili},
\newblock \bibinfo{title}{{Recovering software layers from object oriented
  systems}},
\newblock in: \bibinfo{booktitle}{ENASE 2014 - Proceedings of the 9th
  International Conference on Evaluation of Novel Approaches to Software
  Engineering}, pp. \bibinfo{pages}{1--12}.
\bibitem[{Shatnawi et~al.(2017)Shatnawi, Mili, El~Boussaidi, Boubaker,
  Gu{\'e}h{\'e}neuc, Moha, Privat, and Abdellatif}]{Shatnawi2016}
\bibinfo{author}{A.~Shatnawi}, \bibinfo{author}{H.~Mili},
  \bibinfo{author}{G.~El~Boussaidi}, \bibinfo{author}{A.~Boubaker},
  \bibinfo{author}{Y.-G. Gu{\'e}h{\'e}neuc}, \bibinfo{author}{N.~Moha},
  \bibinfo{author}{J.~Privat}, \bibinfo{author}{M.~Abdellatif},
\newblock \bibinfo{title}{Analyzing program dependencies in java ee
  applications},
\newblock in: \bibinfo{booktitle}{2017 IEEE/ACM 14th International Conference
  on Mining Software Repositories (MSR)}, \bibinfo{organization}{IEEE}, pp.
  \bibinfo{pages}{64--74}.
\bibitem[{Shatnawi et~al.(2018)Shatnawi, Mili, Abdellatif, Boussaidi,
  Gu{\'e}h{\'e}neuc, Moha, and Privat}]{shatnawi2018implement}
\bibinfo{author}{A.~Shatnawi}, \bibinfo{author}{H.~Mili},
  \bibinfo{author}{M.~Abdellatif}, \bibinfo{author}{G.~E. Boussaidi},
  \bibinfo{author}{Y.-G. Gu{\'e}h{\'e}neuc}, \bibinfo{author}{N.~Moha},
  \bibinfo{author}{J.~Privat},
\newblock \bibinfo{title}{How to implement dependencies in server pages of jee
  web applications},
\newblock \bibinfo{journal}{arXiv preprint arXiv:1803.05253}
  (\bibinfo{year}{2018}).
\bibitem[{Kiczales and Lamping(1992)}]{kiczales1992}
\bibinfo{author}{G.~Kiczales}, \bibinfo{author}{J.~Lamping},
\newblock \bibinfo{title}{Issues in the design and specification of class
  libraries},
\newblock in: \bibinfo{booktitle}{Conference Proceedings on Object-oriented
  Programming Systems, Languages, and Applications}, OOPSLA '92,
  \bibinfo{publisher}{ACM}, \bibinfo{address}{New York, NY, USA},
  \bibinfo{year}{1992}, pp. \bibinfo{pages}{435--451}.

\end{thebibliography}
\end{document}